\documentclass[12pt,preprint]{aastex}

\newcommand{\myemail}{jlg@ujaen.es}

\slugcomment{{\sc Accepted for publication in ApJ Letters} }

\shorttitle{Galaxy Gravity}
\shortauthors{Pesta\~{n}a and Eckhardt}

\begin{document}

\title{Cause of Spiral Galaxy Rotation Rates: a Massive Graviton}

\author{Jos\'{e} Luis G. Pesta\~{n}a\altaffilmark{1}}
\affil{Departamento de F\'{\i}sica, Universidad de Ja\'en, Campus Las Lagunillas, 23071 Ja\'en, Espa\~{n}a}

\and

\author{Donald H. Eckhardt}
\affil{Canterbury, NH 03224-0021, USA}

\altaffiltext{1}{Correspondence should be addressed to J. L. G. Pesta\~{n}a (\myemail).}

\begin{abstract}
We attribute the observed rotational velocity profiles of spiral galaxy disks to a combination of Newtonian and exponential gravitational potentials.  We offer a novel, yet perfectly plausible, Lagrangian from which the exponential potential is derived.  
The exponential potential  is defined by two universal constants (corresponding to a coupling constant and a graviton mass) that we determine from our sample of 12 THINGS disks.
All velocity profile fits are excellent.

\end{abstract}


\section{Introduction}

Kepler's third law states that the squares of the periods of any two planets are in the same proportion as the cubes of their mean distances from the Sun; this is explained by the Newtonian inverse-square law of gravitation.  This relationship does not apply to the stars in a spiral galaxy disk because their periods are generally more or less proportional to their distances from the galactic center.  That is, the disk circular velocity $v_c$ profiles of spiral galaxies are observed to be approximately flat, $v_c\approx$ constant. 
 Newtonian gravitation does not explain this, so astrophysicists have been disposed (1) to hypothesize the existence of invisible matter in a halo centered on each galaxy,
and that this dark matter is distributed in a seeming conspiracy to result in the observed motions while retaining the inverse-square law of gravitation, or (2) to modify Newtonian theory.

The Lagrangian for Newtonian gravitation in a centrally symmetric field is
\begin{equation}
{\mathcal L}=-\int_0^{\infty} [(\nabla\psi)^2+8\pi G\rho\psi)]\,d^3r,
\label{eq:LN}
\end{equation}
where $G$ is the gravitational constant, $\psi=\psi(r)$ is the gravitational potential, and $\rho=\rho(r)$ is the mass density.  The solution to the variation $\delta {\mathcal L}=0$ is the Poisson equation,
\begin{equation}
\nabla^2\psi=4\pi G\rho,
\label{peq}
\end{equation}
so the solution to Eq.~\ref{peq} for a point mass $M$ at $r=0$ is the Newtonian potential $\psi_N=-GM/r$, and the consequent centripetal force obeys the inverse-square law.
\citet{Mi:83a,Mi:83b} addressed the disk velocity discrepancy  for galaxies with his Modified Newtonian Dynamics theory, MOND, for which 
 Eq.~\ref{eq:LN} is modified~\citep{BeMi:84} with the replacement
\begin{equation}
(\nabla\psi)^2 \rightarrow a_0^2\,{\mathcal F}\left[(\nabla\psi/a_0)^2\right],
\label{MF}
\end{equation} 
where the acceleration $a_0$ is a universal constant and ${\mathcal F}'(x^2)\approx x$ for $x \ll 1$ and ${\mathcal F}'(x^2)\approx 1$ for $x \gg 1.$  On setting $a_0\approx 10^{-8}$ cm s$^{-2}$, the functional form of ${\mathcal F}$ obliges spiral galaxy disk velocity profiles to be  generally flat, in agreement with observations.
\citet{BM:06} and \citet{MO:11} also devised  non-Newtonian explanations for the flat velocity profiles of spiral galaxies.

\citet{EPF:10} [EPF] considered non-Newtonian gravitation at the scale of galaxy superclusters.  For the Lagrangian,
\begin{equation}
{\mathcal L}=-\int_0^{\infty} [(\nabla\psi)^2+\mu^2\psi^2+8\pi G\rho\psi)]\, d^3r,
\label{eq:LY}
\end{equation}
the solution to $\delta {\mathcal L}=0$ is 
\begin{equation}
(\nabla^2-\mu^2)\psi=4\pi G\rho,
\label{yeq}
\end{equation}
and the solution to Eq.~\ref{yeq} for a point mass $M$ at $r=0$ has the form of a Yukawa potential (see Appendix), 
\begin{equation}
\psi_Y=-GM\exp(-\mu r)/r.
\label{yukpot}
\end{equation}
If $\mu>0$, this results in a Milne universe that is in full accord with cosmological expansion observations of type Ia supernovae but, unlike the explication of \cite{Retal:07},
  it requires neither dark matter ($\Omega_m=0$) nor dark energy ($\Omega_{\Lambda}=0$).
The choice $\mu^{-1}\approx 5$ Mpc then explains the scales of galaxy superclusters and of the fundamental spectrum of the cosmic background radiation; and it explains why neighboring superclusters tend to be aligned in spongiform  ``surfaces''  with vast empty regions between them.

  The graviton mass corresponding to $\mu^{-1}= 5$ Mpc is
$m=\hbar \mu/c =1.3\times10^{-30}$ eV/c$^2$, so EPF conjectured that the mass of the (cosmological) graviton is $m_c\sim 10^{-30}$ eV/c$^2$, and that there could be heavier gravitons  as well, but none that is lighter than $m_c$.  This led us to conjecture that  a heavier (galaxy) $m_g$ graviton is responsible for the flat galaxy velocity curves.  An $m_g$ Yukawa potential cannot explain the flat curves, but
 \cite{eck93} had suggested that an $m_g$ exponential potential combined with a Newtonian potential could provide an explanation (see his Fig. 2).
 However,  his theoretical argument for the existence of an exponential potential was arduous and unconvincing, so we now offer a more substantial argument based on the Lagrangian,
\begin{equation}
{\mathcal L}=-\int  [\mu^{-2}(\nabla^2\psi)^2+ 2 (\nabla\psi)^2+\mu^2\psi^2+8\pi G\rho\psi]\,d^3r.
\label{eq:LE}
\end{equation}
The solution to $\delta {\mathcal L}=0$ is 
\begin{equation}
\mu^{-2}\nabla^4\psi-2\nabla^2\psi+\mu^2\psi=\mu^{-2}(\nabla^2-\mu^2)^2\psi=-4\pi G\rho,
\label{ExE}
\end{equation} 
and  the solution to Eq.~\ref{ExE} for a point mass $M$ at $r=0$ has the form of a exponential potential (see Appendix), 
\begin{equation}
\psi_E =-\gamma GM\,\mu\,\exp(-\mu r),
\label{epot}
\end{equation}
where, for the $m_g$ graviton, $\mu\rightarrow \mu_g=m_gc/\hbar$ and $\gamma\rightarrow \gamma_g$ is a dimensionless constant.

\section{Modeling Galaxy Disks}
At galaxy scales, Yukawa potential due to $m_c$ appears to be Newtonian, so we examine how well the combination of Newtonian and $m_g$ exponential potentials can   reconcile observations of luminosity distributions of spiral galaxies with those of the rotational velocity distributions of the gases and stars in their disks.

Because the centripetal acceleration of a disk object rotating in a circular orbit at velocity $v_c$ and distance $r$ from a galaxy centroid equals the specific force on the object,
\begin{equation}
v_c^2=r(\psi'_N+\psi'_E)\,.
\label{eq:v1}
\end{equation}
We model the Newtonian and $m_g$ potentials of the disk using the double ring technique of \cite{EP:02} [EP], using their Eq.~3 and performing the calculations with Mathematica, Version 8.0.  Adopting the EP [$c_n, R, \sigma$] nomenclature,  we set $c_{-1}=1$ to evaluate $\psi_N$, whereas
 to evaluate $\psi_E$ we use the series expansion of $\exp(-\mu_g R)$ to calculate
 $c_n$ for $n=0,1, \cdots, 10;$ and we then scale up the final result by the factor $\gamma_g$.  The  density distribution $\sigma(R)$ is calculated using all photometrically determined galaxy matter in its central component and disk, plus interstellar gases in the disk, which are determined from the 21-cm radiation of neutral hydrogen - along with a 25 percent by mass admixture of helium.
Our sample of spiral galaxies is a subset of 12 nearby disks selected from the 19 disks in
the HI Nearby Galaxy Survey [THINGS] that were analyzed by \cite{dBetal:08} [dBETAL].
  We excluded seven galaxies {\em a priori} either because they are clearly dominated by noncircular motions (NGC 3031, NGC 3627, NGC 4826 and NGC 4736) or we cannot eliminate that possibility (NGC 2366), or because of the
 the difficulty and uncertainty of their photometric analyses  (NGC 3521 and NGC 3198).
 
 The galaxies in our sample, which were all observed with the same instrument and resolution, extend over a luminosity range of more than three orders of magnitude.  They comprise the best currently available observational data for resolving galaxy rotation profiles that can be used to evaluate diverse gravitational models. 
dBETAL performed a meticulous decomposition of the stellar surface density profiles that  is of utmost importance for testing non-Newtonian dynamical theories; indeed, they even weighted each point of the observed photometric profile by the M/L ratio expected from the observed color gradient as a function of radius. 
   Table~\ref{tbl-dat} lists our sample in order of decreasing rotation velocity. Bear in mind that these  astronomical data, although highly refined, are not perfect:  there always remain lingering uncertainties which, in order of decreasing significance, arise from errors in distance, photometric imprecisions, and sub-estimates of radio observations.

\begin{deluxetable}{lrrrr}
\tablecaption{Model Inputs. Luminosities are in the $3.6\, \mu\mathrm{m}$ band.
\label{tbl-dat}}
\tablewidth{0pt}
\tablehead{
\colhead{Object}&
\colhead{Distance}&
\colhead{Disk Luminosity}&
\colhead{Inner Disk Luminosity}&
\colhead{HI Mass}
\\
&[$\mathrm{Mpc}$]&[$\mathrm{10^9\;L_{\odot}}$]&[$\mathrm{10^9\;L_{\odot}}$]&[$\mathrm{10^8\;M_{\odot}}$]
}
\startdata
NGC 2841&14.10&173.15&21.54&98.91\\
NGC 7331&14.72&398.14&1.66&75.81\\
NGC 6946&5.90&92.61&0.02&37.26\\
NGC 2903&8.90&30.34&1.87&4.63\\
NGC 5055&10.10&169.25&19.77&98.9\\
NGC 3621&6.64&39.76&&67.44\\
NGC 2403&3.47&16.46&0.83&30.42\\
NGC 925&9.20&17.12&&28.37\\
NGC 7793&3.91&9.41&&7.99\\
NGC 2976&3.56&4.60&&1.13\\
IC 2574&4.00&2.24&&13.32\\
DDO 154&4.30&0.82&&3.15
\enddata
\end{deluxetable}

\begin{figure}
\plotone{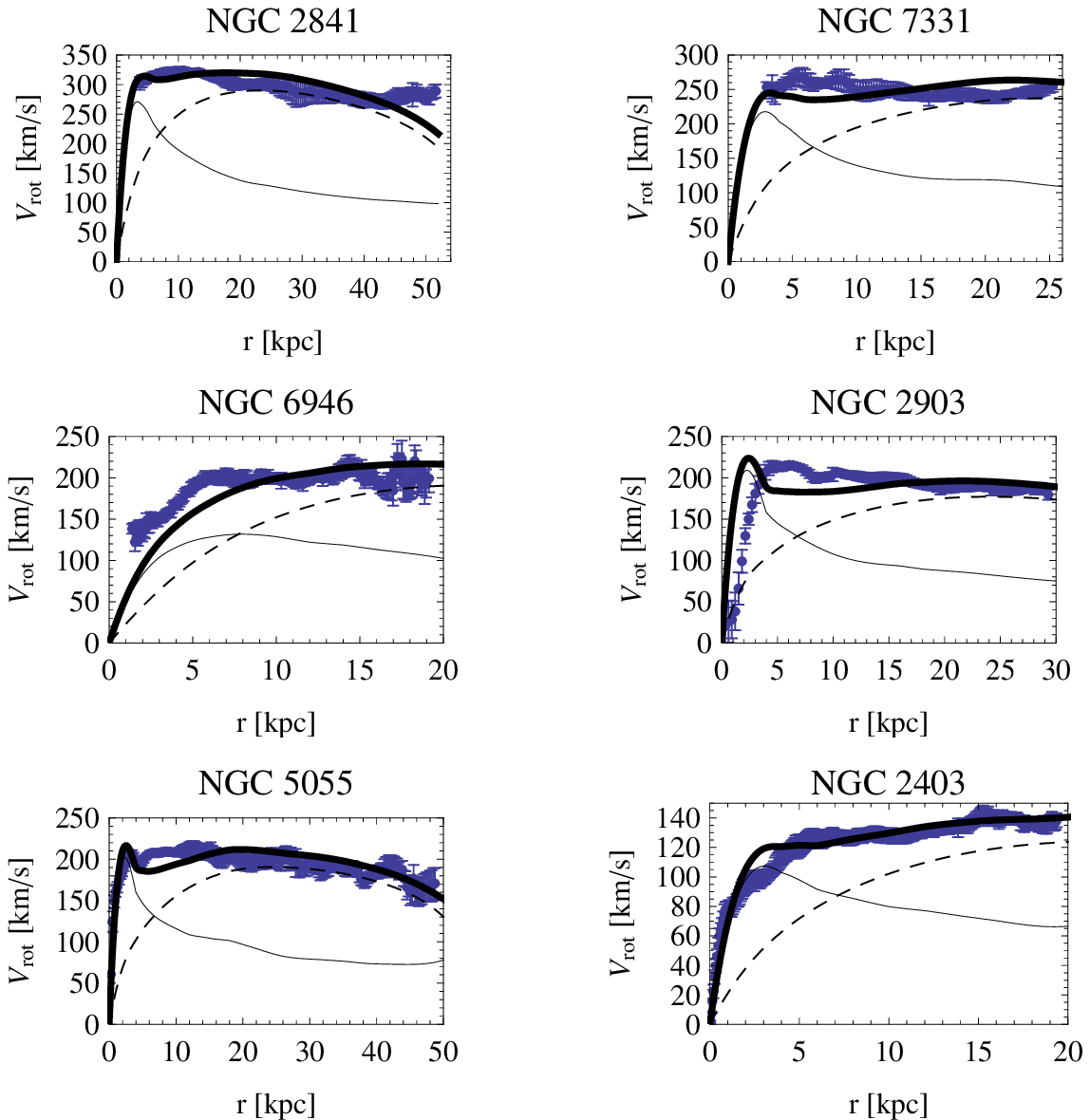}
\caption{Velocity fits for galaxies with two photometric disks. The thick solid black line is $v$. The thin solid lines are the $v$ contributions of the Newtonian potential and the broken lines are those of of the $m_g$ potential.}
\label{fig-res}
\end{figure}

\begin{figure}
\plotone{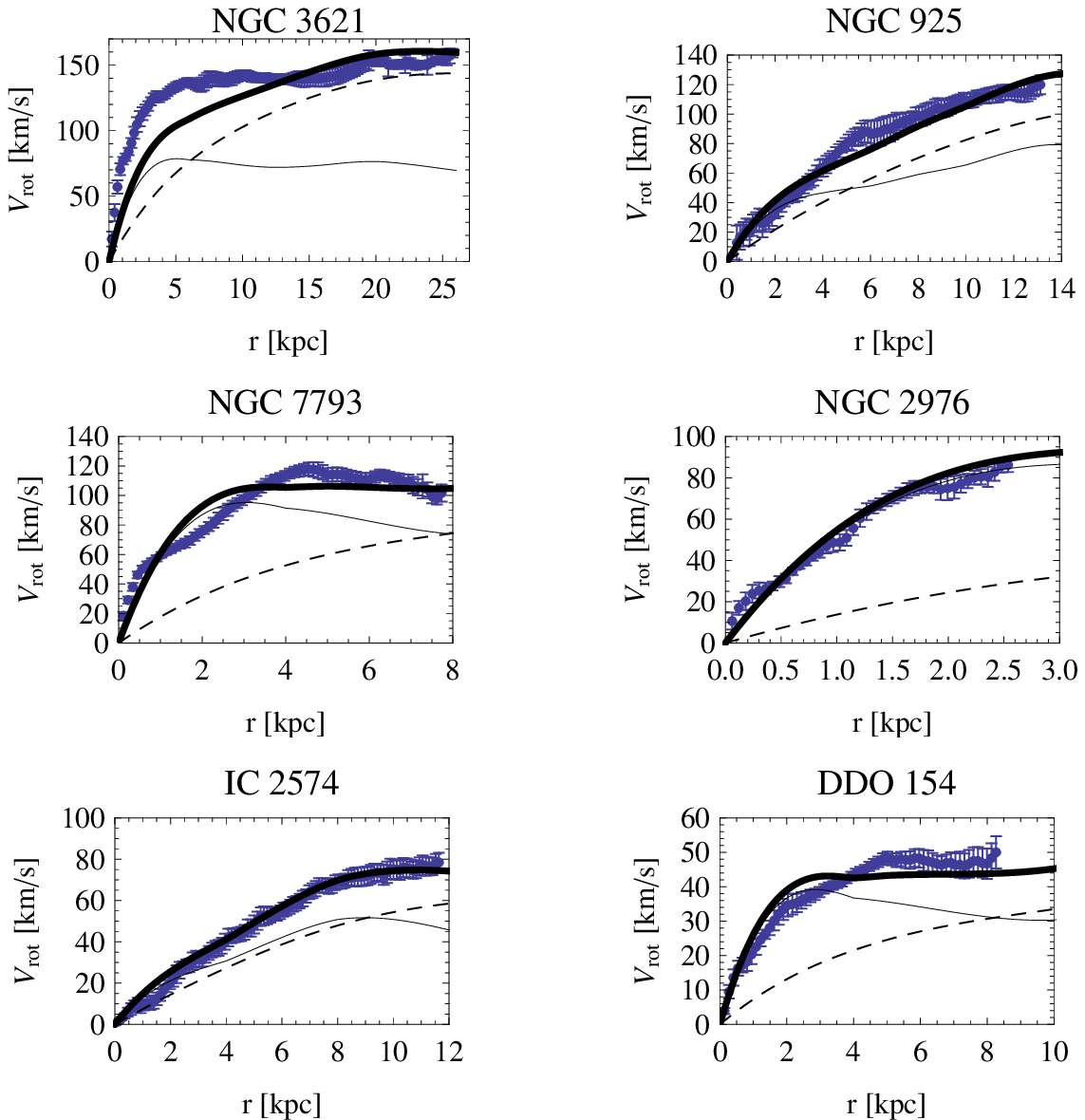}
\caption{Velocity fits for galaxies with just one photometric disk.}
\label{fig-res2}
\end{figure}

To reconcile the velocities, $v_c$, determined by Eq. \ref{eq:v1} with velocities, $v_{obs}$, observed using the 21-cm lines,
we use the observed surface brightness data and color gradients without any adjustments, and the HI densities of dBETAL by applying the double ring technique.  We use this technique for the principal disk and the inner disks (bulges) detected by dBETAL. 
 Then we iteratively adjust the mass-luminosity ratios, M/L for the Spitzer IRAC $3.6\, \mu\mathrm{m}$ band, in each of the 12 dBETAL galaxies' disks, and central components where appropriate, along with the two universal constants, $\gamma_g$ and $\mu_g$, until the $v_c$ and $v_{obs}$ profiles are in close agreement for all galaxies.
 Note that the distances are fixed parameters (the same as those used by dBETAL for their dark halo models) which have uncertainties of $\gtrsim 10\%$.  Allowing these distances to vary within those uncertainties would result in modest improvements to the already excellent fits.  
 We determine that the numerical values of the two fundamental constants  are $\gamma_g=12.5$ and $\mu_g^{-1}=20$ kpc ($m_g=3.2\times 10^{-28}$ eV/c$^2$).  Although we have no formal error estimates for these constants, they are narrowly constrained.
  Table~\ref{tbl-res} lists our dynamically determined values for M/L. There are only one or two free parameters per galaxy, depending on whether (Figure~\ref{fig-res}) or not (Figure~\ref{fig-res2}) the galaxy has an additional internal photometric disk.  By contrast, the dark matter hypothesis requires three free parameters per galaxy, or four parameters for the galaxies with inner disks. 

\begin{deluxetable}{lcr}
\tablecaption{Model Outputs (in solar units).\label{tbl-res}}
\tablewidth{0pt}
\tablehead{
\colhead{Object}&
\colhead{$\mathrm{(M/L)}$}&
\colhead{$\mathrm{(M/L)_{inner}}$}
}
\startdata
NGC 2841&0.10&3.00\\
NGC 7331&0.02&23.00\\
NGC 6946&0.34&0.01\\
NGC 2903&0.15&11.00\\
NGC 5055&0.06&1.00\\
NGC 3621&0.25&\\
NGC 2403&0.50&2.50\\
NGC 925&0.40&\\
NGC 7793&0.70&\\
NGC 2976&0.70&\\
IC 2574&0.70&\\
DDO 154&9.50&
\enddata
\end{deluxetable}

The quality of the adjustment can be assessed by inspecting the curves of Figures~\ref{fig-res} and ~\ref{fig-res2}, and by the plausibility of the dynamically determined ratios M/L. The velocity components in Figure~\ref{fig-res} and ~\ref{fig-res2} combine as root-sum-squares. Each of the 12 panels of the figures shows that the $m_g$ potential component makes up for the discrepancy between the Newtonian component and the observed velocities. The velocity $v_c$ profiles agree very well with the observed velocity $v_{obs}$ profiles, for high as well as low $v_{obs}$, and especially so in view of observational complications and the simplicity of our model. Also, our fits indicate that there is a smooth transition between the small and large disks, with the $m_g$ potential becoming progressively more important as the radial dimensions of the galaxies increase.

For comparisons with Figures~\ref{fig-res} and ~\ref{fig-res2}, dark matter halo fits are shown by dBETAL or \cite{Oetal:08} and, except for NGC 925 and IC 2574, the MOND  fits are shown by   \cite{GFdB:11}.  Note that each dark matter halo fit requires a pair of parameters that is idiosyncratic to the particular galaxy being modeled, $2\times 12=24$ parameters in all, and that the MOND fits require recurrent tweakings of  both the ``universal constant'' $a_0$ and the galaxy distances \citep{Betal:02}.  By contrast, all of our fits are for the same two constants, $\gamma_g$ and $\mu_g$, and the galaxy distances are not modified.

All the dynamically determined M/L ratios in Table~\ref{tbl-res} are unambiguous and, with one exception (DDO 154), they are consistent with stellar population synthesis models.  The same exception was noted by dBETAL in their dark halo modeling; we refer the reader to their paper for a discussion of possible explanations. Also, the inner M/L ratios for NGC 7331 and NGC 2903 appear odd. However, possible explanations could be the presence of a strong dust ring in the inner disk of NGC 7331, and the presence of the bar and a very dense molecular disk in the inner disk of NGC 2903. In fact, both dBETAL and \cite{GFdB:11} also get unrealistic M/L ratios for these two galaxies.

\section{Conclusions and Outlook}

The essential result of this work is that the $m_g$ exponential potential can account for the magnitude of the discrepancy and reproduces the general shapes of the rotation curves of these 12 THINGS spiral galaxies.  Invoking dark matter halos to reproduce the observed shapes requires 24 additional free parameters which have no underlying physical significance; they are chosen just to get satisfactory fits.  And, although MOND purportedly involves only one universal constant, $a_0$, it is actually a parameter that does not remain precisely the same for all fits, and the galaxy distances and the exact functional forms of ${\mathcal F}$ (see Eq.~\ref{MF}) are also sometimes modified.  By contrast, the $m_g$ exponential potential is determined by only two fixed and physically significant parameters: $\mu_g$, which is proportional to the exchange boson mass, and $\gamma_g$, which is proportional to the square of the coupling constant.  The inequality $2<24$ provides a compelling argument for the superiority of the $m_g$ exponential potential over {\em ad hoc} dark matter halos for modeling spiral galaxy rotation profiles.  A similar, although not so lopsided, inequality applies on contrasting the $m_g$ potential fits with those of MOND

 The $m_c$ Yukawa potential is derived from the Eq.~\ref{eq:LY} Lagrangian, and the $m_g$ exponential potential is derived from the Eq.~\ref{eq:LE} Lagrangian.  In the solar system, the $m_c$ potential appears to be Newtonian whereas the $m_g$ potential results in a Rindler type acceleration toward the Sun, $\gamma_g\mu_g^2GM_{\odot}=4\times 10^{-21}$ m s$^{-2}$, which is far below the celestial mechanics detection threshold, $1\times 10^{-15}$ m s$^{-2}$ \citep{i11}.

\appendix

\section{Solutions to Eqs.~\ref{yeq} and \ref{ExE}}

Modify the Fourier integral pair
$$\Phi(k)=\int_0^{\infty} \phi(r)\,\sin kr\,dr,$$
$$\phi(r)=\frac{2}{\pi}\int_0^{\infty} \Phi(k)\,\sin kr\,dk$$
to
$$\frac{\Phi(k)}{k}=\int_0^{\infty} \frac{\phi(r)}{r}\,j_0(kr)\,r^2\,dr,$$
$$\frac{\phi(r)}{r}=\frac{2}{\pi}\int_0^{\infty} \frac{\Phi(k)}{k}\,j_0(kr)\,k^2\,dk,$$
where the spherical Bessel function $j_0(kr)$ is simply
$$j_0(kr)=\frac{\sin kr}{kr}.$$
Next make the replacements $\Phi(k)/k\rightarrow\Psi(k)$ and $\phi(r)/r\rightarrow\psi(r).$  This gives a Fourier-Bessel  integral pair:
\begin{equation}
\Psi(k)=T[\psi(r)]=\int_0^{\infty} \psi(r)\,j_0(kr)\,r^2\,dr,
\end{equation}
\begin{equation}
\psi(r)=T^{-1}[\Psi(k)]=\frac{2}{\pi}\int_0^{\infty} \Psi(k)\,j_0(kr)\,k^2\,dk
=\frac{1}{i \pi  r}\int_{-\infty}^{\infty} \Psi(k)\,\exp ikr\,k\,dk.
\end{equation}
Then
$$\nabla^2j_0(kr)=-k^2j_0(kr),$$
$$\nabla^2\psi(r)=T^{-1}[-k^2\,\Psi(k)],$$ and
$$T[\nabla^2\psi(r)]=-k^2\,\Psi(k).$$
Suppose that $\rho=\rho(r)=\rho_0$ for $r\le r_0$, where $kr_0\ll 1$ (so $j_0(kr)\approx 1$), and that $\rho(r)=0$ for $r>r_0$; then
$$T[4\pi G\rho]\approx G\rho_0\int_0^{r_0}4\pi r^2\,dr=GM.$$
This becomes an equality if $M$ is a point mass, and $$T[(\nabla^2-\mu^2)\psi-4\pi G\rho]=-(k^2+\mu^2)\Psi(k)-GM=0,$$
or
\begin{equation}
\Psi(k)=-\,\frac{GM}{k^2+\mu^2},
\label{Psiy}
\end{equation}
$$\psi=T^{-1}[\Psi(k)]=-\frac{GM}{i\pi r}\int_{-\infty}^{\infty}\frac{k\, \exp ikr}{(k+i\mu)(k-i\mu)}\,dk.$$
Contour integration in the upper half plane about the pole $k=i\mu$ then gives the Yukawa potential of Eq.~\ref{yukpot},
$\psi=\psi_Y$. Following the same approach to solve Eq.~\ref{ExE}, the only modification is that Eq.~\ref{Psiy} is replaced by
\begin{equation}
\hat{\Psi}(k)=-\,\frac{\mu^2GM}{(k^2+\mu^2)^2}.\label{Psie}
\end{equation}
Rather than perform a contour integration about a double pole, it is simpler just to note that
$$-\mu^2\,\frac{\partial \Psi(k)}{\partial (\mu^2)}=\hat{\Psi}(k),$$
so the corresponding potential is modified to
$$\hat{\psi}=-\mu^2\,\frac{\partial \psi_Y}{\partial (\mu^2)}=-\frac{\mu}{2}\,\frac{\partial \psi_Y}{\partial \mu}=-\frac{GM}{2}\,\mu\,\exp(-\mu r).$$
But $G$ here is a generic constant anyway, with the same dimensions as the gravitational constant $G$ used in 
Eq.~\ref{yukpot}, so we simply replace $G/2$ by $\gamma G$, where $\gamma$ is a dimensionless constant that takes into account the $1/2$ factor and the differences between the coupling constants for the different potentials.  The solution to Eq.~\ref{ExE} is then the exponential potential given by Eq.~\ref{epot}.
The $\mu$ is not subsumed into $\gamma$ because $\mu$ has the dimension of inverse distance, as does $1/r$ in the Yukawa potential.

\end{document}